\RequirePackage{lineno}
\documentclass[prb,twocolumn,amsmath,amssymb,showpacs]{revtex4}
\usepackage{graphicx}
\usepackage{natbib}
\usepackage{bm}
\usepackage[usenames]{color}

\definecolor{red}{RGB}{1,1,1}

\begin{document}

\title{Comment on ``Gap opening in graphene by shear strain''}

\author{Ashwin Ramasubramaniam}
\email{ashwin@engin.umass.edu}
\affiliation{Department of Mechanical and Industrial Engineering, University of Massachusetts Amherst, Amherst MA 01003}

\date{\small\today}

\def\linenumberfont{\normalfont\small\sffamily}

\begin{abstract}

G. Cocco, E. Cadelano, and L. Colombo [Phys.~Rev.~B {\bf 81}, 241412(R) (2010)] have suggested that combinations of shear and uniaxial strain can be used to open a band gap in graphene at much lower levels of strain than with the application of unaxial strain alone. They employed a \emph{unit} cell of graphene in their studies and applied the Cauchy-Born rule to model external strain. Consequently, an important aspect of the mechanical behavior of membranes, namely buckling and wrinkling under external strain, and the attendant coupling with electronic structure was ignored in their analysis. Upon doing so, the apparent band gap that appears in the range of 15--20\% shear strain under the Cauchy-Born assumption is shown to vanish. The gapless spectrum of graphene is found to persist under large shear strains as well as large combinations of shear and uniaxial strain. 
\end{abstract}


\maketitle


In a recent article, Cocco \emph{et al.} \cite{Cocco} have suggested that a combinations of shear and uniaxial strains in the range of 12--17\% can be used to open a band gap in graphene. Previous studies \cite{Castro Neto} indicate that uniaxial strains as high as 23\% along the zigzag direction are required to open a band gap in graphene; uniaxial strains along the armchair direction do not induce a band gap. Since strains $\sim 23 \%$ are close to the failure strength of graphene,\cite{Kysar} Cocco \emph{et al.}'s proposal, which relies on lower levels of strain, presents a more practical approach for strain-induced band gap opening in graphene.

In the following, I show that Cocco \emph{et al.}'s suggestion, while interesting, is unlikely to induce band gaps at any reasonable (i.e.~well below failure) levels of strain in a graphene sheet. In particular, I show that neither shear strains as high as 20\% nor combinations of shear and uniaxial strains as high as 14\% open a gap in graphene. These results rely on the key observation that shearing a graphene sheet inevitable leads to out-of-plane deformation of the sheet. Such bucking/wrinkling behavior is well known from the theory of membranes\cite{buckling} and, moreover, has been observed experimentally in graphene membranes.\cite{Lau}  Cocco \emph{et al.}'s studies that employ only a unit cell of graphene are incapable of capturing such wrinkling and buckling of a graphene sheet. Consequently, there are important fundamental differences between their results and the ones I discuss next.

The simulation cell consists of a 49.43\AA$\,\times\,$49.43\AA~ graphene sheet (20 unit cells along each primitive lattice vector) consisting of 800 C atoms [Fig.\,1]. Periodic boundary conditions are applied along the cell vectors. Further computational details are provided at the end of this Comment. An in-plane shear is applied to the sheet via a deformation gradient
\footnote{The deformation gradient $\bm F$ acts on a vector $\bm X$ in the reference configuration and transforms it to a vector $\bm x= \bm F \bm X$ in the current configuration; simultaneously, a vector $\bm q$ in the reciprocal lattice transforms as $\bm F^{-T} \bm q$.}
\begin{equation}
\bm F = \left[ 
\begin{array}{ccc}
1 & \gamma & 0 \\
0 & 1 & 0\\
0 & 0 & 1 
\end{array}
\right].
\end{equation}
The corresponding Lagrange strain tensor 
\footnote{The small strain tensor of linear elasticity $\bm \epsilon$ is obtained from the Lagrange strain tensor in the limit $\gamma \ll 1$. Since shear strains $\sim$15--20\% are certainly not small, we will have no further use for $\bm \epsilon$. Instead, the deformation of the simulation cell and its Brillouin zone is completely characterized by $\bm F$. For making connection with Ref.\,\onlinecite{Cocco}, one may loosely interpret the shear strain $\zeta$ in their notation as $\gamma/2$ here.}
is
\begin{equation}
\bm E = \frac{1}{2} [\bm F^{T} \bm F - \bm I]
=
\frac{1}{2}\left[ 
\begin{array}{ccc}
0 & \gamma & 0 \\
\gamma & \gamma^{2} & 0\\
0 & 0 & 0
\end{array}
\right].
\end{equation}
Two scenarios for the morphology of the deformed sheet are considered: (A) no relaxation of atomic positions is permitted, the displacement of every atom being slaved to the macroscopic deformation $\bm F$  (Cauchy-Born rule), (B) relaxation of atomic positions is permitted.%
\footnote{Cell vectors and angles are not relaxed for now.} 
Scenario A is essentially identical to that of Ref.\,\onlinecite{Cocco} apart from the number of unit cells in the calculation. The resulting morphologies for both scenarios for $\gamma = 0.4$ are displayed in Fig.\,1(a). Clearly, large out-of-plane displacements occur at this level of shear. Buckled morphologies at various levels of shear strain are displayed in Fig.\,1(b); unsurprisingly, larger applied shears lead to larger buckling amplitudes. The relative energies of the buckled morphologies with respect to the flat ones are displayed in Table 1. By buckling out-of-plane, the graphene sheet greatly reduces the stored elastic energy; this is especially apparent at large deformations.

\begin{figure*}
\begin{center}
\includegraphics[width=\textwidth]{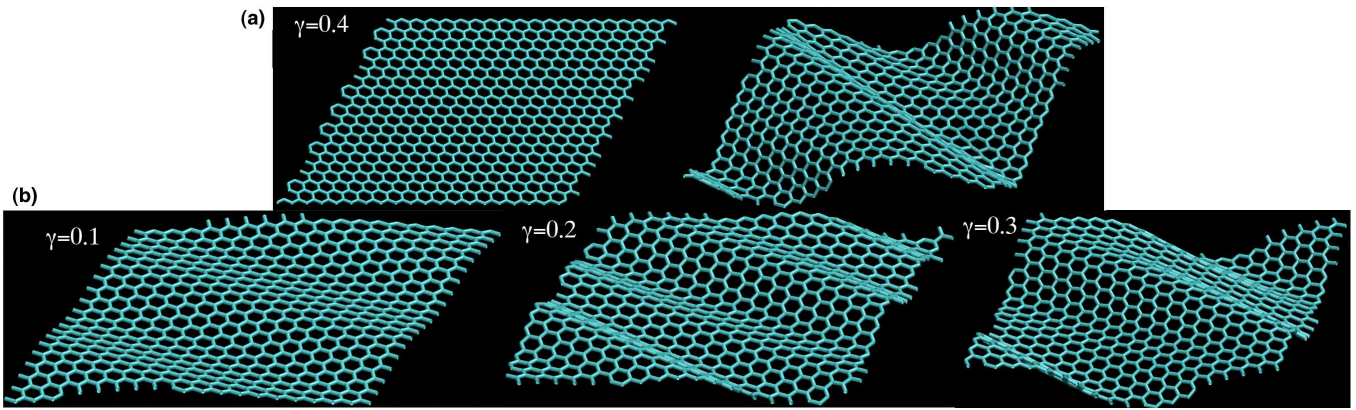}
\end{center}
\caption{(a) Morphology of the graphene sheet subjected to a shear $\gamma=0.4$ with (left) and without (right) relaxation of atoms in the simulation cell. (b) Morphologies of buckled sheets at various levels of shear employed in this study.}
\end{figure*}

\begin{table}[h!]
\caption{Relative energy $\Delta E$ of buckled sheet with respect to unbuckled sheet for various applied shears $\gamma$.}
\centering
\begin{tabular}{|c| c|}
\hline
$\gamma$  & $\Delta E$ [eV/atom] \\
\hline
0.1 &   -0.04 \\
0.2 & -0.22 \\
0.3 & -0.67 \\
0.4 & -1.10  \\
\hline
\end{tabular}
\end{table}

Next, we consider the electronic structure of the flat and buckled sheets as a function of applied shear $\gamma$. Fig. 2(a) displays the density of states (DOS) as a function of applied shear under Scenario A. As seen from this figure, there is no band gap opening up to $\gamma=0.3$. At $\gamma=0.4$, we open up a gap of $\sim 1.28$ eV. The DOS is qualitatively similar (albeit quantitatively different) from that of Ref.\,\onlinecite{Cocco}. In agreement with Ref.\,\onlinecite{Cocco}, we see that a band gap opens up in the range $\gamma=0.3-0.4$ . Fig.\,2(b) displays the DOS for the buckled sheets at the same levels of applied shear $\gamma$. It is immediately apparent that there is \emph{no longer any evidence of a band gap} at $\gamma=0.4$. Clearly, there is an intimate connection between the morphology of the graphene sheet and its electronic structure. Constraining the sheet in an artificial configuration dictated by the Cauchy-Born rule keeps it from attaining physically relevant low energy states, which in turn is seen to lead to an artificial band gap at higher levels of shear.

\begin{figure*}
\begin{center}
\includegraphics[width=\textwidth]{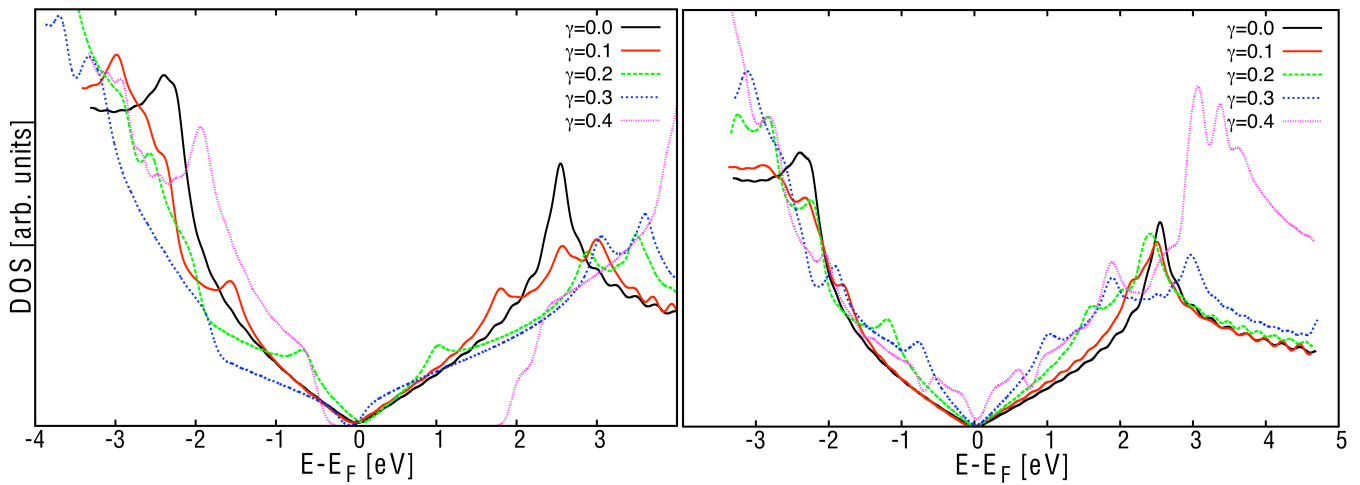}
\end{center}
\caption{Density of states (DOS) of graphene sheets (a) with and (b) without atomic relaxation as a function of applied shear $\gamma$. From (a) it is clear that a band gap opens up in the range $\gamma=0.3-0.4$ if atomic relaxation is not permitted. However, as seen from (b) if the sheet is allowed to attain a lower energy state via out-of-plane buckling, there is no evidence of a gap at $\gamma=0.4$.}
\end{figure*}

Finally for completeness, let us consider the case of combined shear and uniaxial strain along the armchair direction; application of strain along the zigzag direction in conjunction with shear does not induce a gap as was shown in Ref.\,\onlinecite{Cocco} and verified independently in this work . The applied deformation gradient is chosen to be of the form
\begin{equation}
\bm F = \left[ 
\begin{array}{ccc}
1 & \gamma & 0 \\
0 & 1+\epsilon & 0\\
0 & 0 & 1 
\end{array}
\right],
\end{equation}
with $\epsilon=0.14$ and $\gamma=0.28$. 
\footnote{It is worth noting that this strain state is close to the maximum that can be sustained by the sheet, as may be ascertained by computing the eigenvalues of the Lagrange strain tensor. For  $\epsilon=0.15$ and $\gamma=0.3$, which corresponds roughly to the case of combined uniaxial and shear strain of $\zeta=15\%$ in Ref.\,\onlinecite{Cocco}, the sheet begins to develop vacancies and holes.}
The morphology of the relaxed sheet is shown in Fig.\,3(a). As seen from Fig.\,3(b), we obtain a band gap of $\sim 0.29$ eV in the DOS under the Cauchy-Born assumption, which once again disappears upon allowing for atomic relaxation. 

\begin{figure*}
\begin{center}
\includegraphics[width=0.9\textwidth]{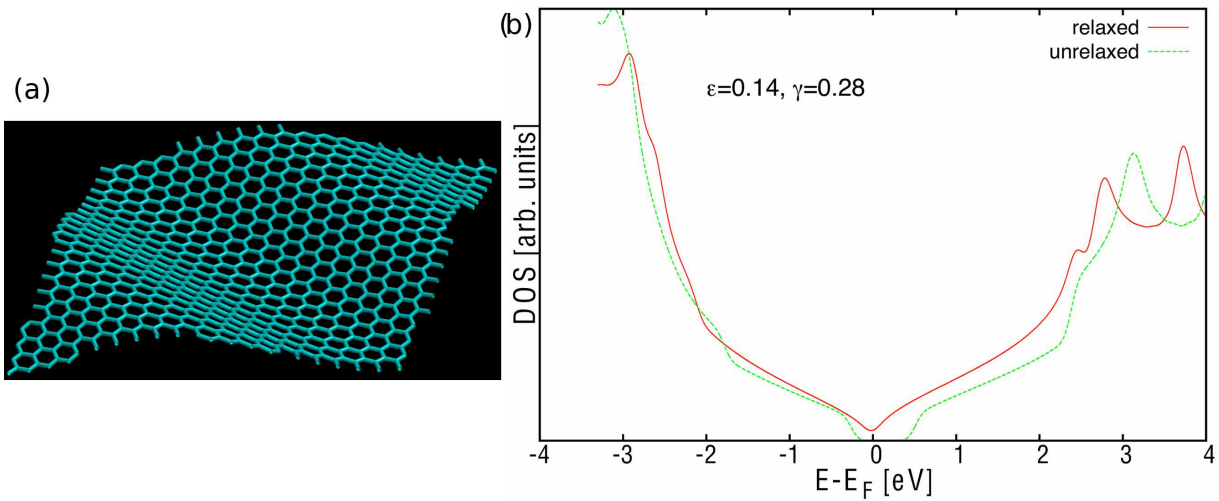}
\end{center}
\caption{Combined shear $\gamma=0.28$ and uniaxial strain $\epsilon=0.14$ along the armchair direction. (a) Morphology of relaxed sheet. (b) Density of states for unrelaxed and relaxed sheet. As seen, the band gap of $\sim 0.29$ eV obtained in the unrelaxed case vanishes upon allowing for atomic relaxation.}
\end{figure*}

In conclusion, it appears that the results of Cocco \emph{et al.} for gap-opening in graphene under shear, while strictly correct for a unit cell of graphene, are unlikely to be applicable in practice to a graphene sheet of reasonable physical dimensions. The results in this Comment clearly indicate that it is crucial to account for buckling of the graphene sheet and the intimate coupling between the morphology of the sheet and its electronic structure. Upon doing so, it would appear that opening a band gap in graphene once again requires extremely high strains, close to the failure strain of the material. Thus, the results in this Comment, in conjunction with those drawn in the work of Pereira \emph{et al.},\cite{Castro Neto} seem to reinforce the notion that the gapless spectrum of graphene is indeed quite robust under mechanical deformation.

\emph{Computational details}: Electronic structure calculations were performed with the DFTB$^{+}$ code \cite{DFTB} employing the Slater-Koster parameters generated by Elstner \emph{et al.} \cite{SKparam}  A $4\times 4\times 1$ Monkhorst-Pack mesh is used for Brillouin zone sampling. A tolerance of $10^{-5}$ Ha is used for the self-consistent charge cycles. All atomic positions are relaxed with a force tolerance of 0.01 Ha/Bohr. The DOS is obtained by smearing the eigenvalue spectrum with Gaussians of variance $\sigma^{2}=0.005$. 

Useful discussions with D.~Naveh are gratefully acknowledged.

\end{document}